\input phyzzx.tex
\hoffset=10mm
\vsize=26cm
\titlepage 
\title{{\bf Yang's Quantized Space-time Algebra and Holographic Hypothesis} }
\author{Sho TANAKA\footnote{*}{Em.~Professor of Kyoto University and 
Associate Member of Institute of Qauntum Science, Nihon University.
E-mail: stanaka@yukawa.kyoto-u.ac.jp}}
\address{Kurodani 33-4, Sakyo-ku, Kyoto 606-8331, Japan}
\abstract{The present-day significance of Yang's quantized space-time algebra (YST) is pointed out from the holographic viewpoint. One finds that the D-dimensional YST and its modified version (MYST) have the background symmetry SO(D+1,1) and SO(D,2), which are well known to underlie the dS/CFT and AdS/CFT correspondences, respectively. This fact suggests a new possibility of dS/YST and AdS/MYST in parallel with dS/CFT and AdS/CFT, respectively. The spatial components of quantized space-time and momentum operators of YST have discrete eigenvalues and their respective minimums, $a$ and $1/R$, without contradiction to Lorentz-covariance. With respect to MYST, the spatial components of space-time operators and the time component (energy) of momentum operators have discrete eigenvalues and their respective minimums, $a$ and $1/R$, in contrast to YST. This discrete structure of YST and MYST, which CFT lacks entirely, may provide a theoretical ground for the unified regularization or cutoff of ultraviolet and infrared divergences familiar in the UV/IR connection in the holographic hypothesis.}

\endpage

\chapter{Introduction}
In the previous paper\ref\Tanaka (referred to as I, hereafter), the present author attempted second quantization of the matrix model of $D_0$ branes, that is, quantum mechanics of $D_0$ branes proposed by  Banks, Fischer, Shenker and Susskind,\Ref\Bf\bfss\ by introducing second-quantized D field defined on quantized space-time early proposed by Snyder\ref\snyder and immediately after by Yang.\Ref\Y\yang 

\def\Tanaka{S.~Tanaka, ``Space-time quantization and nonlocal field theory--Relativistic second quantization of matrix model," hep-th/0002001.}
\def\bfss{T.~Banks, W.~Fischler, S.~H.~Shenker and L.~Susskind, ``M Theory As A Matrix Model: A Conjecture," Phys.~Rev.~{\bf D55} (1997), 5112, hep-th/9610043.}
\def\snyder{H.~S.~Snyder, Phys.~Rev.~{\bf 71} (1947), 38; {\bf 72} (1947),
68.}
\def\yang{C.~N.~Yang, Phys.Rev. {\bf 72} (1947), 874; Proc. of International 
Conf.~on Elementary  Particles, 1965 Kyoto, pp 322-323.}
\def\sw{Among a number of works with respect to the holographic hypothesis or the UV/IR connection, it is worthwhile, for the present argument, to notice:\nextline Susskind and Witten, ``The Holographic Bound in Anti-de Sitter Space," hep-th/9805114;\nextline L.~Dyson, J.~Lindesay and L.~Susskind, ``Is There Really a de Sitter/CFT Duality," JHEP {\bf 0208} (2002) 045, hep-th/0202163.} %

The purpose of the paper I was two-fold: First, it makes creation and annihilation of $D_0$ branes possible beyond the limitation of the matrix model\refmark{\Bf} as N-body quantum-mechanical system of $D_0$ branes. Second, by introducing second quantized D field, we tried to rewrite the matrix model in a Lorentz-covariant way. 

Indeed, it should be noted here that through this procedure; (i) the noncommutative $N\times N$ position operators (observables) of $N$ quantum-mechanical $D_0$ branes defined on a single light-cone time noncovariantly in the matrix model disappear, and (ii) they are replaced by second quantized D field defined on the manifestly covariant quantized space-time, which is irrelevant to the position coordinates of individual $D_0$ branes.
In this attempt, we further found that Yang's quantized space-time is appropriate to describe the matrix model, as compared with the original Snyder's quantized space-time.   

\def\dfr{S.~Doplicher, K.~Fredenhagen and J.E.~Roberts, ``Space-time Quantization induced by Classical Gravity,"  Phys.~Lett. {\bf B331} (1994), 39; \nextline  
D.~Minic, ``M-THEORY AND DEFORMATION QUANTIZATION," hep-th/9909022; \nextline 
L.~Smolin, ``M theory as a matrix extension of Chern-Simons theory," Nucl.~Phys. {\bf B591} (2000), 227, hep-th/0002009; \nextline
 B.-D.~D{\" o}rfel, ``A Lie-algebra model for a noncommutative space time geometry," hep-th/020416.}
In this paper, the present author wishes further to notice that Yang's quantized space-time has the distinct characteristics among many recent attempts of noncommutative space-time algebra,\ref\dfr\ which aim to overcome the nonrelativistic aspects inherent to the widely known noncommutative space-time with $c$-number $\theta^{\mu \nu}$. 

First of all, one should note that the noncommutative algebra satisfied by D-dimensional Yang's quantized space-time (YST), has a background symmetry SO(D+1,1) common to D-dimensional Euclidean conformal algebra (CFT) and (D+1)-dimensional de Sitter algebra, which is widely known to underlie the dS/CFT correspondence in the holographic hypothesis.\Ref\C\dS\ Indeed, as will be shown in the next section, one finds that D-dimensional YST is derived through the dimensional contraction of SO(D+1,1) algebra, by choosing two dimensions with space-like metric signature as extra dimensions. 

On the other hand, it is well known that D-dimensional Euclidean conformal algebra and (D+1)-dimensional de Sitter algebra are derived through the dimensional contraction of the above SO(D+1,1) algebra, by choosing two dimensions with opposite metric signature and one dimension with space-like metric signature as extra dimensions, respectively. This fact strongly suggests a new possibility of ${dS}_{D+1}/{YST}_D$ correspondence in parallel with the well-known $dS_{D+1}/{CFT}_D$ correspondence\refmark{\C}. 

At this point, it is very interesting to consider the modified version of YST, i.e., MYST which has a background symmetry SO(D,2) common to D-\break dimensional conformal algebra and (D+1)-dimensional anti-de Sitter algebra, underlying the AdS/CFT correspondence. In this case, one should notice the possibility of  ${AdS}_{D+1}/{MYST}_D$ correspondence in parallel with the well-known \break ${AdS}_{D+1}/{CFT}_D$ correspondence. In fact, as will be shown in section 3, D-\break dimensional MYST is derived through the dimensional contraction of SO(D,2) algebra, by choosing two dimensions with opposite metric signatures as extra dimensions, in close connection with D-dimensional conformal algebra.    
\def\dS{ R.~Bousso, ``Holography in General Space-times," JHEP {\bf 9906} (1999) 028, hep-th/9906022; \nextline E.~Witten, ``Quantum Gravity In De Sitter Space," hep-th/0106109; \nextline 
A.~Strominger, ``The dS/CFT Correspondence," JHEP {\bf 0110} (2001) 034, hep-th/0106113; \nextline R.~Bousso, A.~Maloney and A.~Strominger, ``Conformal Vacua and Entropy in de Sitter Space," Phys.~Rev. {\bf D65} (2002) 104039, hep-th/0112218. }

In addition, one should notice that the spatial components of quantized space-time and momentum operators of YST have discrete eigenvalues and their respective minimums, $a$ and $1/R$, without contradiction to Lorentz-covariance. With respect to MYST, the spatial components of space-time operators and the time component (energy) of momentum operators have discrete eigenvalues and their respective minimums, $a$ and $1/R$, in contrast to YST. 

This discrete structure of YST and MYST, which CFT lacks entirely, may provide a theoretical ground for the unified regularization or cutoff of ultraviolet and infrared divergences familiar in the UV/IR connection in the holographic hypothesis.\Ref\Sw\sw\  The reason why such a discreteness is not explicit in the conformal algebra which has the same symmetry SO(D+1,1) or SO(D,2) with Yang's space-time algebra as stated above, will be also clarified.

In section 2, the algebraic structure of Yang's quantized space-time (YST) is clarified in connection with the dS/CFT correspondence and in particular discrete structure of eigenvalues of generators is examined in detail, keeping in mind the so-called UV/IR connection in the holographic hypothesis. Section 3 is devoted to the consideration of a modified version of YST in connection with AdS/CFT correspondence, by use of the arguments on the original YST given in connection with dS/CFT correspondence in section 2. Concluding remarks are given in section 4.  

\chapter{Yang's quantized space-time algebra and dS/CFT correspondence}

 D-dimensional Yang's quantized space-time is given through the dimensional contraction of SO(D+1,1) algebra with generators $\Sigma_{MN}$ in the following way. First of all, D-dimensional space-time and momentum
operators, $X_\mu$ and $P_\mu$, with $\mu =1,2,...,D,$ where $\mu=D$ is time-like, are defined by
$$
     X_\mu \equiv a\ \Sigma_{\mu a}
\eqn\eqA
$$
$$
     P_\mu \equiv 1/R \ \Sigma_{\mu b},   
\eqn\eqB
$$
together with D-dimensional angular momentum operator $M_{\mu \nu}$
$$
   M_{\mu \nu} \equiv \Sigma_{\mu \nu}
\eqn\eqC
$$ 
and the so-called reciprocity operator
$$
    N\equiv a/R\ \Sigma_{ab}.
\eqn\eqD
$$
In the above expression, both $a$ in \eqA\ and $R$ in \eqB\ must be fundamental constants of Yang's quantized space-time, as will be seen below.  $\Sigma_{MN}$ is defined by
$$
 \Sigma_{MN}  \equiv i\ (y_M \partder{}{y_N}-y_N\partder{}{y_M})
\eqn\eqE
$$
in terms of (D+2)-dimensional Yang's parameters $y_M$  with $M,N= (\mu,a,b)$. They satisfy
$$
             - y_0^2 + y_1^2 + ... + y_{D-1}^2 + y_a^2 + y_b^2 = R^2,
\eqn\eqF
$$
where $y_0 =-i y_D$ and $M = a, b$ denote two extra dimensions with space-like metric signature.
These operators  $( X_\mu, P_\mu, M_{\mu \nu}, N )$ defined above constitute Yang's quantized space-time algebra, YST:
$$
   [ X_\mu, X_\nu ] = - i a^2 M_{\mu \nu}
\eqn\eqG
$$
$$
   [ P_\mu, P_\nu ] = - i/ R^2\ M_{\mu \nu}
\eqn\eqH
$$
$$
      [X_\mu, P\nu ] = - i   N \delta_{\mu \nu}
\eqn\eqI 
$$
$$
     [ N, X_\mu ] = - i a^2  P_\mu
\eqn\eqJ
$$
$$
      [ N, P_\mu ] = i / R^2\ X_\mu,
\eqn\eqK
$$
omitting the obvious relations concerning $M_{\mu \nu}$. 

At this point, it is important to notice the following simple fact that 
$\Sigma_{MN}$ with $M, N$ being the same metric signature have discrete eigenvalues  and those with $M, N$ being opposite metric signature have continuous eigenvalues, as was shown explicitly in I. 

Indeed, this fact tells us that the spatial components of space-time and momentum operators, $X_i$ and $P_i$ with $i = 1,2,...,D-1$, have discrete eigenvalues and their respective minimums, $a$ and $1/R$, while the time-components, $X_0$ and $P_0$, have continuous eigenvalues, as seen in the following expressions:
$$
    X_i = a\ \Sigma_{ia} = a\ i\ (y_i \partder{}{y_a}-y_a\partder{}{y_i})
\eqn\eqL
$$
$$
    X_0  = a\ \Sigma_{0a} =  a\ i\ (y_0 \partder{}{y_a} + y_a\partder{}{y_0})
\eqn\eqM
$$  
and 
$$
   P_i = 1/R\  \Sigma_{ib} = 1/R\  i\ (y_i \partder{}{y_b}-y_b\partder{}{y_i})
\eqn\eqN
$$
$$
   P_0  = 1/R\ \Sigma_{0b} =  1/R\ i\ (y_0 \partder{}{y_b} + y_b\partder{}{y_0}).
\eqn\eqO
$$  

Let us turn our attention to the conformal algebra, which has been recently much noticed in the holographic hypothesis.\refmark{\C} \refmark{\Sw} It is well known that D-dimensional Euclidean conformal algebra is derived through the dimensional contraction of the same SO(D+1,1) algebra considered above. In fact, the generators ${\hat P}_u, {\hat K}_u, {\hat M}_{u v}$ and $\hat D $ are given in the following form,
$$
       ( {\hat P}_u + {\hat K}_u )/2 = \Sigma_{u 0}
\eqn\eqP
$$
$$
       ( {\hat P}_u - {\hat K}_u )/2= \Sigma_{u b}
\eqn\eqQ
$$
$$
       \hat D = \Sigma_{b0}
\eqn\eqR
$$
and
$$
          {\hat M}_{uv} = \Sigma_{u v}.
\eqn\eqS
$$

From the above expressions, one finds that two components with opposite metric signature, $M=0$ and $b$ are chosen as extra dimensions transverse to the D-dimensional Euclidean components, $u = (i, a)$ with $i=1,2,...,D-1$. As a result, the momentum operator ${\hat P}_u$ and the conformal boost operator ${\hat K}_u$ defined through \eqP\ and \eqQ, respectively, turn out to be commutative among themselves, that is, 
$$
[ {\hat P}_u, {\hat P}_v] = [{\hat K}_u {\hat K}_v] =0 
\eqn\eqSS
$$
and have continuous eigenvalues, in contrast to YST. The dilatation operator $\hat D$ in \eqR\ has also continuous eigenvalues and satisfies
$$
\eqalign{        &[ {\hat P}_u,  {\hat D} ] = +i\ {\hat P}_u  \cr 
          &[{\hat K}_u, {\hat D}] = -i\ {\hat K}_u
}
\eqn\eqT
$$
and 
$$
        [{\hat K}_u, {\hat P}_v ] = 2 i\ ( {\delta}_{uv} {\hat D} + {\hat M}_{uv} ).
\eqn\eqU
$$
${\hat P}_u - {\hat K}_u$ and D-dimensional angular momentum operators ${\hat M}_{uv}$ have discrete eigenvalues.
 
Now let us examine (D+1)-dimensional de Sitter algebra, which is also related to the SO(D+1,1) algebra through the dimensional contraction and widely discussed in connection with the above conformal algebra in the so-called dS/CFT correspondence.\refmark{\C} The generators $( {\check P}_\alpha, {\check M}_{\alpha \beta})$ are defined in the following way;
$$
    {\check P}_\alpha = 1/R\ \Sigma_{\alpha b}
\eqn\eqV
$$
and 
$$
       {\check M}_{\alpha \beta} = {\Sigma}_{\alpha \beta},
\eqn\eqW
$$
which satisfy
$$  
   [ {\check P}_\alpha, {\check P}_\beta] = - i/ R^2\ {\check M}_{\alpha \beta}.\eqn\eqX
$$

In this case, one finds that the space-like component $M=b$ is chosen as an extra dimension transverse to the (D+1)-dimensional components $\alpha = (a, \mu)$ of de Sitter space. The spatial components of momentum operator ${\check P}_u$ with $u =1,2,...,D-1,a$ have discrete eigenvalues, like those of YST, as seen from the relations
$$
\eqalign{
     &{\check P}_\mu = P_\mu \cr    
      &{\check P}_a = 1/a\ N
}
\eqn\eqY
$$
with $\mu =0,1,2,...,D-1$.

In the preceding arguments, we have examined the discreteness of eigenvalues of generators in YST, conformal algebra and de Sitter algebra, all of which are derived as the result of different contractions of the same SO(D+1,1) algebra, whose $(D+2)(D+1)/2$ generators $\Sigma_{MN}$ have the definite discreteness, that is, either discrete or continuous, as mentioned before. Consequently, it turns out that the discreteness of generators of each algebra simply depends on the respective contractions, where the relevant generator is defined as a certain linear combination of $\Sigma_{MN}$. One can see its typical example in conformal algebra, where both generators ${\hat P}_u$ and ${\hat K}_u$ defined through \eqP\ and \eqQ\ are expressed as superposition of ${\Sigma}_{u0}$ and ${\Sigma}_{ub}$ with continuous and discrete eigenvalues, respectively, and thus their discrete aspect entirely disappears.

In addition, one notices that in both conformal algebra and de Sitter (or anti-de Sitter, see section 3) algebra, space-time coordinates are entirely out of algebraic consideration and assumed as continuous number from the beginning. This treatment leads us naturally to the conventional local field theory based on the continuous space-time, but at the same time to the danger of various divergences, as seen in the argument of the holographic hypothesis.\refmark{\Sw} 

By contrast, in YST, which has the background symmetry SO(D+1,1) common to the former two algebras, the spatial components of quantized space-time and momentum operators have discrete eigenvalues and their respective minimums, $a$ and $1/R$, as was pointed out above. 

 This consideration strongly suggests a new possibility that Yang's space-time algebra (YST) plays an important role in the background of the dS/CFT correspondence or more directly in the dS/YST correspondence. Indeed, they may provide a theoretical ground for the unified regularization or cutoff of ultraviolet and infrared divergences which has been recently much discussed in the so-called UV/IR connection or holographic bound.\refmark{\Sw}

\chapter{Modified version of YST and AdS/CFT correspondence}

At this point, it is very interesting to consider a modified version of Yang's quantized space-time algebra, i.e., MYST which is obtained by choosing extra dimension $b$ to be time-like and taking $R$ replaced with $iR$ in the preceding arguments in section 2.  Eq. \eqF\ is now replaced by
$$
             - y_0^2 + y_1^2 + ... + y_{D-1}^2 + y_a^2 - y_{0'}^2 = - R^2,
\eqn\eqZ
$$
with $y_{0'} = -i y_b.$

In this case, D-dimensional MYST becomes to have the background symmetry SO(D,2) with generators ${\Sigma'}_{MN}$ being $\Sigma_{M N}$ \eqE\ with $M, N=b$ and $D$ chosen to be time-like. This symmetry is well known to be common to D-dimensional conformal algebra and (D+1)-dimensional anti-de Sitter algebra, underlying the well-known ${AdS}_{D+1}/{CFT}_{D}$ correspondence.

In fact, D-dimensional MYST $(X'_\mu, P'_\mu, M'_{\mu, \nu}, N')$ is given by
$$
             X'_\mu \equiv a\ \Sigma'_{\mu a}
\eqn\eqAA
$$
$$             P'_\mu \equiv -i / R\  \Sigma'_{\mu b}
\eqn\eqAB
$$
$$
            M'_{\mu \nu} \equiv  {\Sigma'}_{\mu \nu}
\eqn\eqAC
$$
and 
$$
                 N' \equiv -i a / R\ \Sigma'_{a b},
\eqn\eqAD
$$
with $\mu , \nu = 1, 2, ...,D,$ corresponding to \eqA, \eqB, \eqC\ and \eqD.

In this case, (D+1)-dimensional anti-de Sitter algebra $( {\check P'}_\alpha, {\check M'}_{\alpha \beta})$ is given by 
$$
    {\check P'}_\alpha = - i/R\ {\Sigma'}_{\alpha b}
\eqn\eqAV
$$
and 
$$
       {\check M}_{\alpha \beta} = {\Sigma'}_{\alpha \beta}
\eqn\eqAW
$$
with $\alpha, \beta = (a, \mu),$ corresponding to \eqV\ and \eqW.

On the other hand, D-dimensional conformal algebra (${\hat P'}_\mu, {\hat K'}_\mu, {\hat M'}_{\mu \nu}, \hat D' $) is given in the form closely related to the above MYST $(X'_\mu, P'_\mu, M'_{\mu, \nu}, N')$: 
$$
       ( {\hat P'}_\mu + {\hat K'}_\mu )/2 =  \Sigma'_{\mu a}\ (= 1/a\ X'_\mu)
\eqn\eqAE
$$
$$
       ( {\hat P'}_\mu - {\hat K'}_\mu )/2= -i \Sigma'_{\mu b}\ (= R\ P'_\mu)
\eqn\eqAF
$$
$$
      {\hat M'}_{\mu \nu} \equiv  {\Sigma'}_{\mu \nu}\ ( = M'_{\mu \nu})    
\eqn\eqAG
$$
and
$$
        \hat D' = -i \Sigma'_{ab}\  ( = R/a\  N').
\eqn\eqAH
$$

Here, one should notice that with respect to the modified Yang's quantized space-time algebra (MYST), the spatial components of the space-time operators \eqAA\ have discrete eigenvalues and minimum $a$, in accord with YST discussed in section 2. However, the spatial components of momentum operators \eqAB, together with the reciprocity operator \eqAD, tend to have continuous eigenvalues, and the time component (energy) to have discrete eigenvalues and minimum $1/R$, in contrast to YST. 

This implies that MYST may play an important role in the background of the AdS/CFT correspondence or more directly in the AdS/MYST correspondence, entirely in parallel with the argument on YST given in the preceding section in connection with the dS/CFT or dS/YST correspondence. 

It should be noted again that, as seen from Eqs. \eqAE\ and \eqAF, the discrete structure of eigenvalues of ${\hat P'}_\mu$ and ${\hat K'}_\mu$ of conformal algebra entirely disappears in accord with the argument in section 2.

\chapter{Concluding Remarks}

In the preceding arguments, we have clarified that the D-dimensional Yang's quantized space-time algebra (YST) and its modified version (MYST) have the background symmetry SO(D+1,1) and SO(D,2), which underlie the dS/CFT and AdS/CFT correspondences and possibly the dS/YST and AdS/MYST correspondences, respectively. 

Furthermore, we found that their respective generators have the proper discreteness of eigenvalues, which CFT lacks entirely. That is, the spatial components of quantized space-time operators in both YST and MYST have discrete eigenvalues and their minimum $a$, while the spatial components of the momentum operators in YST and the time component (energy) in MYST have discrete eigenvalues and their minimum $1/R$. This discrete structure proper to YST and MYST may provide a theoretical ground for the unified regularization or cutoff of ultraviolet and infrared divergences familiar in the UV/IR connection or holographic bound in the dS/CFT or AdS/CFT correspondence.     

It should be however further studied in the future how the existence of minimums or discreteness of eigenvalues of space-time and momentum operators as seen in Yang's noncommutative space-time algebra may lead us actually to cutoff of ultraviolet and infrared divergences, in place of the conventional cutoff theory on the continuous space-time. 
\def\FL{H.~Firouzjahi and F.~Leblond, ``The clash between de Sitter and anti-de Sitter space," hep-th/0209248.}

Finally, with respect to the so-called background-dependence of space-time, it should be noted that SO(D+1,1) in YST or SO(D,2) in MYST considered in the background of Yang's quantized space-time algebra is nothing but a set of Killing vector fields realized on the (D+2)-dimensional pseudo-Euclidean space, \eqF\ or \eqZ, which is assumed for (D+2)-dimensional Yang's parameter space. At this point, it is interesting to conjecture that the pseudo-Euclidean frame may be dynamically changeable\ref\FL between SO(D,2) and SO(D+1,1), and it might be a kind of {\it local reference frame} chosen at any given point in (D+2)-dimensional Yang's parameter space, on the analogy of the familiar {\it local Lorentz frame} in general theory of relativity.

\ack{The present author would like to thank S.~YAHIKOZAWA for the valuable conversation and critical comments.}
\refout
\end